\documentclass[11pt,twoside]{article}

\usepackage{asp2004}
\usepackage{epsf}
\usepackage{psfig}
\usepackage{lscape}
\usepackage{graphicx}

\newcommand{\msunyr}{\ensuremath{\mathit{M}_{\odot}{\rm yr}^{-1}}}   

\newcommand{\mdot}{\ensuremath{\dot{M}}}                             
\newcommand{\vinf}{\ensuremath{v_{\infty}}}                          

\begin{document}
\title{Radio observations of candidate magnetic O stars}    
\author{R.S. Schnerr$^1$, K.L.J. Rygl$^1$, A.J. van der Horst$^1$, T.A. Oosterloo$^2$, J.C.A. Miller-Jones$^1$ 
and H.F. Henrichs$^1$}   
\affil{$^1$ Sterrenkundig Instituut "Anton Pannekoek", Kruislaan 403,
 1098 SJ Amsterdam, the Netherlands\\
$^2$ ASTRON, Oude Hoogeveensedijk 4, 7991 PD Dwingeloo, the
 Netherlands}    

\begin{abstract} 
A number of O stars are suspected to have (weak) magnetic fields because of the observed cyclical variability in their UV wind-lines. However, direct detections of these magnetic fields with optical spectropolarimetry have proven to be very difficult. We have searched for non-thermal radio emission, which would be a strong indication for the presence of a magnetic field. Of our 5 selected candidate magnetic O stars, 3 are detected: $\xi$ Per, which we find to have a non-thermal spectrum, and $\lambda$ Cep and $\alpha$ Cam which show a thermal spectrum. We also find that the fluxes are lower than the expected free-free (thermal) contribution of the stellar wind. This is in agreement with recent findings that the mass-loss rates from O stars using H$\alpha$ are overestimated because of clumping in the inner part of the stellar wind.
\end{abstract}

\section{Introduction}
Free-free radiation of ionised stellar winds gives rise to a thermal radio spectrum which can be calculated using:
{\footnotesize
\begin{equation}
\label{eq:thermal}
S_\nu = 7.26 \left[ \frac{\nu}{10\,\mathrm{GHz}} \right]^{0.6} \left[ \frac{T_e}{10^{4}\,\mathrm{K}} \right]^{0.1}
          \left[ \frac{\mdot}{10^{-6}\,\msunyr} \right]^{4/3} 
       \left[\frac{\mu_e \vinf}{100\, \mathrm{km s^{-1}}} \right]^{-4/3} \left[\frac{D}{\mathrm{kpc}} \right]^{-2} \mathrm{mJy}
\end{equation}}
\noindent \citep{scuderi:1998,panagia:1975,wright:1975}, where $D$ is the distance, \mdot~is the mass-loss rate, $T_e$ the electron temperature, $\mu_e$ the mean atomic weight per electron, \vinf\ the terminal wind velocity and $\nu$ the frequency. However, about 30\% of O stars are found to show non-thermal radio emission \citep[e.g.][]{bieging:1989}. This is characterised by a flatter-than-thermal spectrum, or $\alpha < 0.6$, with $S_\nu \propto \nu^\alpha$. The most likely mechanism for non-thermal emission in the radio regime is synchrotron radiation, which requires a magnetic field. 
This was supported by the detection of non-thermal radio emission from the strongly magnetic Ap/Bp stars \citep{drake:1987}.

The majority of O stars show cyclical variability in their UV wind-lines \citep{kaper:1996}. The timescale of this variability is of the order of the estimated rotation period (days) and could be well explained by (weak) co-rotating magnetic fields, which affect the base of the wind.

\section{Observations \& Results}
We have observed 5 O stars with known cyclical wind-line variability with the Westerbork Synthesis Radio Telescope (WSRT) at 1.4 GHz (21 cm) between  September and November 2005. In addition, 10 Lac was observed at 4.8 GHz (6 cm, WSRT) and $\xi$ Per at 8.5 GHz (3.6 cm, VLA archive). Combined with measurements from the literature, the range in frequencies allows a determination of the spectral index $\alpha$. 

\begin{figure}[!t]
\includegraphics[width=\linewidth]{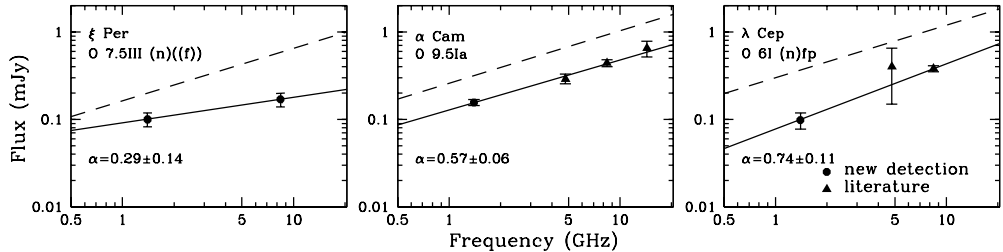}
\caption{Radio data of $\xi$\,Per, $\lambda$\,Cep and $\alpha$\,Cam. Circles: our measurements; triangles: literature values. Solid line: least-squares fit with a powerlaw of index $\alpha$; dashed line: predicted free-free flux (powerlaw with $\alpha=0.6$).}
\label{plot}
\end{figure}

We have detected 3 candidate magnetic O stars at radio wavelengths. $\xi$ Per shows a non-thermal spectrum; $\alpha$ Cam and $\lambda$ Cep show a thermal spectrum (see Fig.~\ref{plot}). The detection of non-thermal emission in $\xi$ Per strengthens the case that the UV line variability observed in this star is caused by a magnetic field.
Van Loo (2005) and \citet{vanloo:2005b} conclude from numerical simulations that both a magnetic field and a binary companion are required to explain non-thermal radio emission, but no binary companion is known for $\xi$ Per.

In general the observed radio flux is lower than predicted by Eq.~\ref{eq:thermal}. This confirms recent results by \citet{fullerton:2006} and \citet{puls:2006astro_ph} that mass-loss rates derived from free-free radio emission are significantly lower than those derived from H$\alpha$ modeling, which is a signature of enhanced clumping in the inner part of the stellar wind.

{\acknowledgements We thank the WSRT staff and ASTRON. This research made use of the VLA archives.}

\end{document}